\documentclass[12pt]{article}
\usepackage{here,amsfonts,pstricks,pst-node,
pst-plot,pst-coil,epsfig,psfig}
\newbox\sample
\newif\ifproofmode
\newif\ifsymindex
\global\symindexfalse
\newwrite\inx
\def\indsyma#1#2{\ifproofmode\marginpar{$\scriptstyle#1$}\fi%
\ifx#2\empty\write\inx{$\noexpand#1$,\space\thepage}%
\write\inx{\string\newline}\else%
\write\inx{$\noexpand#1$,\space#2,\space\thepage}%
\write\inx{\string\newline}\fi\ignorespaces}%
\def\indsym#1#2{\ifsymindex%
\ifproofmode\marginpar{$\scriptstyle#1$}\fi%
\ifx#2\empty\write\inx{\string\item \space$\noexpand#1$,\space\thepage}%
\else%
\write\inx{\string\item \space$\noexpand#1$,\space#2,\space\thepage}%
\fi\ignorespaces\fi}%
\newskip\dangerskipb
\newskip\dangerskip
\def\hang{\hangindent\dangerskip}

\def\s#1{{\cal #1}}

\def\lag{\left\langle}
\def\rag{\right\rangle}

\def\pairt#1#2{\lag #1, #2\rag}

\def\proof{\noindent{\it Proof\/}.\enspace}

\def\remark{\bigskip\noindent{\bf Remark:}\enspace}

\font\manual=manfnt at 12pt
\def\danbend{{\manual\char127}}
\def\ddatanger{\medbreak\begingroup\clubpenalty=10000
 \def\par{\endgraf\endgroup\medbreak} \noindent\hang\hangafter=-2
 \hbox to0pt{\hskip-3.5pc\danbend\kern1pt%
\danbend\hfill}}

\def\dobdownarrow{\mathop{\vbox{\kern2pt \hbox{$\Big\downarrow$}\kern-16.5pt
                          \nointerlineskip\hbox{$\Big\downarrow$}}}}

\def\lrightarrow{\hbox to 25pt{\rightarrowfill}}

\def\supexp{exp(m,n,p)=m^{m^{m^{\cdot^{\cdot^{\cdot^{m^{p}}}}}}}
\vbox{\hbox{$\Big\}\scriptstyle n$}\kern0pt}}

\def\supexpo#1#2#3{#1^{#1^{\cdot^{\cdot^{\cdot^{#1^{#2}}}}}}
\vbox{\hbox{$\Big\}\scriptstyle #3$}\kern0pt}}

\def\sqr#1#2{{\vcenter{\hrule height .#2pt
         \hbox{\vrule width.#2pt height#1pt \kern#1pt
             \vrule width.#2pt}
         \hrule height.#2pt}}}

\def\bigsquare{\mathchoice\sqr76\sqr76\sqr{2.1}3\sqr{1.5}3}

\def\lag{\langle}
\def\rag{\rangle}

\def\co{\colon}

\newskip\bogcentering \bogcentering= 0pt plus 1000pt minus 1000pt 

\def\matth{\mathsurround=0pt}
\def\fakrightarrowfill{$\matth \mathord- \mkern-6mu
  \cleaders\hbox{$\mkern-2mu \mathord- \mkern-2mu$}\hfill
 \mkern-6mu \mathord\rightarrow$}

\def\fakoverrightarrow#1{\vbox{\ialign{##\crcr
  \fakrightarrowfill\crcr\noalign{\kern-1pt\nointerlineskip}
 $\hfil\displaystyle{#1}\hfil$\crcr}}}

\def\cases#1{\left\{\,\vcenter{\normalbaselines\matth
  \ialign{$##\hfil$&\quad##\hfil\crcr#1\crcr}}\right.}

\newif\ifdtatp

\def\displaty{%
\global \dtatptrue \openup \jot \matth \everycr{\noalign{\ifdtatp \global 
\dtatpfalse \vskip -\lineskiplimit \vskip \normallineskiplimit \else 
\penalty \interdisplaylinepenalty \fi }}}

\def\displaylignes#1{\displaty
   \halign{\hbox to\displaywidth{$\displaystyle##$}\crcr
   #1\crcr}}
\def\eqaligneno#1{\displaty \tabskip=\bogcentering
 \halign to\displaywidth{\hfil$\displaystyle{##}$\tabskip=0pt
 &$\displaystyle{{}##}$\hfil\tabskip=\bogcentering
 &\llap{$##$}\tabskip=0pt\crcr
 #1\crcr}}
\def\leqaligneno#1{\displaty \tabskip=\bogcentering
 \halign to\displaywidth{\hfil$\displaystyle{##}$\tabskip=0pt
 &$\displaystyle{{}##}$\hfil\tabskip=\bogcentering
 &\kern-\displaywidth\rlap{$##$}\tabskip=\displaywidthpt\crcr
 #1\crcr}}
\def\ligne{\hbox to\hsize}
\newdimen\nouvpagewidth
\newdimen\offwidth
\newdimen\lawidthoui
\nouvpagewidth=\lawidthoui
\def\kboxit#1{\vbox{\hrule\hbox{\vrule\kern3pt
              \vbox{\kern3pt#1\kern3pt}\kern3pt\vrule}\hrule}}

\def\kboxitb#1{\vbox{\hrule\hbox{\vrule\kern3pt
              \vbox{\kern3pt#1\kern3pt}\kern3pt\vrule}\hrule}}

\def\laboxaround#1{
\aboxaround{\hbox to\hsize{\hfill\box2\hfill}}{#1}
}

\def\boxar#1#2{
\aboxaround{\hbox to\hsize{\hfill#1\hfill}}{#2}
}

\def\aboxaround#1#2{
\setbox4=\vbox{\hsize #2\noindent\strut#1\strut}
\kboxitb{\box4}}

\def\kframeit#1{\vbox{\hrule\hbox{\vrule\kern5pt
              \vbox{\kern5pt#1\kern5pt}\kern5pt\vrule}\hrule}}
\newskip\savnormalbaselineskip
\newskip\savnormallineskip
\newdimen\savnormallineskiplimit

\newtheorem{thm}{Theorem}[section]

\def\reals{\mathbb{R}}
\def\affreal{\reals}
\def\ptb{}
\def\mdeg{m}
\def\ndeg{m}
\def\pdeg{p}
\def\qdeg{q}
\def\natnums{\mathbb{N}}
\def\mapdef#1#2#3{#1\co #2\rightarrow #3}

\begin{document}

\title{Fast and Simple Methods For Computing  Control Points}
\author{$\mathrm{Jean~Gallier}^*$ \qquad\qquad $\mathrm{Weiqing~Gu}^{**}$ \\ 
\\
$\>^*$Department of Computer and Information Science\\
University of Pennsylvania\\
Philadelphia, PA 19104, USA \\
\\
$\>^{**}$Mathematics Department\\
Harvey Mudd College\\
1250 N. Dartmouth Ave\\
Claremont, CA 91711\\ \\
{\tt jean@saul.cis.upenn.edu}\\
{\tt gu@math.hmc.edu}
}

%\date{16 Aug. 1996}
\maketitle

\newcommand{\binomalal}[2]{\left(\begin{array}{c}#1\\#2\end{array} \right)}
\vspace{0.3cm}
\noindent
{\bf Abstract.}
The purpose of this paper is to present simple
and fast methods for computing
control points for polynomial curves and polynomial surfaces 
given explicitly in terms
of polynomials (written as sums of monomials).
We give recurrence formulae w.r.t. arbitrary affine frames.
As a corollary, it is amusing that we can also give
closed-form expressions in the case of the frame
$(r, s)$ for curves, and the frame $((1, 0, 0), (0, 1, 0), (0, 0, 1)$ for
surfaces.
Our methods have the same low polynomial (time and space) complexity
as the other best known algorithms,
and are very easy to implement.

\vfill\eject
\section{Introduction}
\label{sec1}
Polynomial curves and surfaces are used extensively in geometric
modeling and computer aided geometric design (CAGD) in
particular (see Ramshaw \cite{Ramshaw87}, Farin \cite{Farin93,Farin95}, 
Hoschek and Lasser \cite{Hoschek}, or Piegl and Tiller \cite{Piegl}). 
One of the main reasons
why polynomial curves and surfaces are used so extensively in CAGD,
is that there is a very powerful and versatile algorithm to recursively
approximate a curve or a surface using repeated affine interpolation,
the {\it de Casteljau algorithm\/}. However, the de Casteljau algorithm
applies to curves and surfaces only if they are defined in terms
of {\it control points\/}. There are situations where
a curve or a surface is defined explicitly in terms of polynomials.
For example, the following polynomials define a surface
known as the Enneper surface:
$$\eqaligneno{
x(u, v) &= u - \frac{u^3}{3} + uv^2\cr
y(u, v) &= v - \frac{v^3}{3} + u^2v\cr
z(u, v) &= u^2 - v^2.\cr
}$$
Thus, the problem of computing
control points from polynomials (defined
as sums of monomials)  arises. 
If control points can be computed quickly from polynomials,
all the tools available
in CAGD for drawing curves and surfaces can be applied.
This could be very useful in problems 
where a curve of a surface is obtained analytically
in terms of polynomials or rational functions, for example,
problems involving generalizations of Voronoi diagrams,
or motion planning problems. When dealing which such problems,
it is often necessary to decide whether curves segments or surface
patches  intersect or not. 
As is well known (for example, see \cite{Farin93,Farin95}),
there are effective methods based on subdivision
(exploiting the fact that a B\'ezier curve or surface patch
is contained within the convex hull of its control points)
for deciding whether B\'ezier curve segments or surface patches intersect.
Simple and fast methods for computing control points might also
be also useful to teach say, Math students, to learn computational tools
for drawing interesting curves and surfaces.
Now, it turns out that the problem of computing control points
can be viewed as a change of polynomial basis, more specifically
as a change of basis from the monomial basis to bases
of Bernstein polynomials. 
Algorithms for performing such changes of basis have been
given by Piegl and Tiller \cite{Piegl}. More general algorithms
for performing changes of bases between progressive bases
and P\'olya bases are presented in 
Goldman and Barry \cite{GoldBarry} and
Lodha and Goldman \cite{LodhaGold}. 
These algorithms compute  certain 
triangles or tetrahedras whose nodes are labeled with
certain multisets, and are generalizations of the de Casteljau
and the de Boor algorithm. 
In this paper, we present  alternate and more direct methods 
for computing control points from polynomial definitions
(in monomial form) 
that run in the same low time complexity as the above algorithms
($O(m^2)$ for curves of degree $\mdeg$, 
$O(\pdeg^2\qdeg^2)$ for rectangular surfaces of bidegree $\pairt{\pdeg}{\qdeg}$,
and $O(m^3)$ for triangular surfaces of total degree $\mdeg$).
Our algorithms are not as general as those of 
Goldman and Barry \cite{GoldBarry} and
Lodha and Goldman \cite{LodhaGold}, but they are more direct
and very easy to implement.

\medskip
The paper is organized as follows.
In section \ref{sec2}, we review briefly the relationship between
polynomial definitions and control points. We begin with the polarization
of polynomials in one or two variables, and then we show how polynomial
curves and surfaces are completely determined by sets of control points.
In the case of surfaces, depending on the mode of
polarization, we get two kinds of surfaces, bipolynomial surfaces
(or rectangular patches) and total degree surfaces (or triangular patches).
Efficient methods for computing control points are given in the next  
three section:
polynomial curves in section \ref{sec5},
bipolynomial surfaces in section \ref{sec3}, and
polynomial total degree surfaces in section \ref{sec4}.
Some examples are given in
section \ref{sec6}.

\section{Control Points}
\label{sec2}
\subsection{Polynomial Curves}
The deep reason why polynomial curves and surfaces can
be handled in terms of control points is that
polynomials in one or several variables can be {\it polarized\/}.
This means that every polynomial function arises from a unique
symmetric multiaffine map.
A detailed treatment of this approach can be found in
Ramshaw \cite{Ramshaw87}, Farin \cite{Farin93,Farin95}, 
Hoschek and Lasser \cite{Hoschek}, 
or Gallier \cite{Gallbook}. We simply review what is needed
to explain our algorithms.

\medskip
Recall that a map $\mapdef{f}{\reals^d}{\reals^n}$ is {\it affine\/} if
$$f((1 - \lambda) a + \lambda b) = (1 - \lambda) f(a) + \lambda f(b),$$
for all $a, b\in \reals^d$, and all $\lambda\in\reals$.
A map $\mapdef{f}{\underbrace{\reals^d\times \cdots\times \reals^d}_{m}}{\reals^n}$ 
is {\it multiaffine\/} if it is affine in each of its arguments, and
a map  $\mapdef{f}{\underbrace{\reals^d\times \cdots\times \reals^d}_{m}}{\reals^n}$ 
is {\it symmetric\/} if it does
not depend on the order of its arguments, i.e.,
$f(a_{\pi(1)},\ldots, a_{\pi(m)}) = f(a_1,\ldots,a_m)$
for all $a_1, \ldots, a_m$, and all permutations $\pi$.
We also say that a map
$\mapdef{f}{\affreal^{\pdeg}\times \affreal^{\qdeg}}{\reals^d}$
is {\it $\pairt{\pdeg}{\qdeg}$-symmetric\/} if
it is symmetric separately in its first $\pdeg$ arguments
and in its last  $\qdeg$ arguments.

\medskip
Let us first treat the case of polynomials in one variable, which corresponds
to the case of curves. 
Given a (plane) polynomial curve $\mapdef{F}{\affreal}{\affreal^2}$ of degree
$\mdeg$, 
$$\eqaligneno{
x(t) &= F_1(t),\cr
y(t) &= F_2(t),\cr
}$$
where $F_1(t)$ and $F_2(t)$ are polynomials of degree $\leq \mdeg$,
it turns out that
$\mapdef{F}{\affreal}{\affreal^2}$ comes from a unique symmetric 
multiaffine map $\mapdef{f}{\affreal^\mdeg}{\affreal^2}$, 
the {\it polar form of $F$\/}, such that
$$F(t) = f(\underbrace{t, \ldots, t}_\mdeg),
\quad\hbox{for all $t\in\affreal$.}$$

\medskip
Furthermore, given any interval $(r, s)$ (affine frame), the map
$\mapdef{f}{\affreal^\mdeg}{\affreal^2}$ is determined by
the sequence $(b_0,\ldots, b_\mdeg)$ of $\mdeg + 1$ {\it  control points\/} 
$$b_i = f(\underbrace{r, \ldots, r}_{\mdeg - i}, 
\underbrace{s, \ldots, s}_i),$$
where $0\leq i \leq \mdeg$.
Using linearity, in order to polarize a polynomial of one variable $t$,
it is enough to polarize a monomial $t^k$. Since there are
$\binomalal{m}{k}$ terms in the sum
$$  \sum_{I \subseteq \{1, \ldots, \mdeg\}\atop
            \scriptstyle |I| = k}
         \biggl(\prod_{i\in I} t_i\biggr),$$
the polar form $f_k^m(t_1,\ldots,t_{\mdeg})$
of the monomial $t^k$ with respect to the degree $m$
(where $k\leq m$) is given by
$$f_k^m(t_1,\ldots,t_{\mdeg}) = 
\frac{1}{\binomalal{m}{k}}\,  \sum_{I \subseteq \{1, \ldots, \mdeg\}\atop
            \scriptstyle |I| = k}
         \biggl(\prod_{i\in I} t_i\biggr).$$

\subsection{Polynomial Surfaces Polarization}
Given a polynomial surface $\mapdef{F}{\affreal^2}{\reals^3}$, 
there are two natural ways to polarize the polynomials
defining $F$.

\medskip
The first way  is to
polarize {\it separately\/} in $u$ and $v$.
If $\pdeg$ is the highest degree in $u$ and $\qdeg$ 
is the highest degree in $v$,
we get a unique $\pairt{\pdeg}{\qdeg}$-symmetric
degree $(\pdeg + \qdeg)$ multiaffine map
$$\mapdef{f}{\affreal^{\pdeg}\times \affreal^{\qdeg}}{\reals^3},$$
such that
$$F(u,v) = f(\underbrace{u,\ldots,u}_{\pdeg}; 
\underbrace{v,\ldots,v}_{\qdeg}).$$

\medskip
We get what is traditionally called
a {\it  tensor product surface\/}, or as we prefer to call it,
a {\it bipolynomial surface of bidegree $\pairt{\pdeg}{\qdeg}$\/} 
(or a {\it rectangular surface patch\/}).

\medskip
The second way to polarize is to treat the variables
$u$ and $v$ {\it as a whole\/}.
This way,  if $F$ is a polynomial surface such that the maximum total degree
of the monomials is  $\mdeg$,
we get a unique symmetric degree $\mdeg$ multiaffine map
$$\mapdef{f}{(\affreal^2)^{\mdeg}}{\reals^3},$$
such that
$$F(u,v) = f(\underbrace{(u, v),\ldots,(u, v)}_{\mdeg}).$$

\medskip
We get what is called a {\it total degree surface\/}
(or a {\it triangular surface patch\/}).

\medskip
Using linearity, it is clear that
all we have to do is to polarize a monomial $u^{h}v^{k}$.

\medskip
It is easily verified that the unique $\pairt{\pdeg}{\qdeg}$-symmetric 
multiaffine polar form of degree $\pdeg + \qdeg$
$$f^{\pdeg, \qdeg}_{h, k}(u_{1},\ldots, u_{p}; v_{1},\ldots,v_{q})$$
of the monomial $u^{h}v^{k}$ is given by

$$f^{\pdeg, \qdeg}_{h, k}(u_{1},\ldots, u_{p}; v_{1},\ldots,v_{q})
= \frac{1}{\binomalal{p}{h} \binomalal{q}{k}}\,
\sum_{{I\subseteq\{1,\ldots,p\}, |I|=h}\atop {J\subseteq\{1,\ldots, q\}, |J|= k}}
\left(\prod_{i \in I} u_{i}\right)\left(\prod_{j \in J}v_{j}\right).
$$

\medskip
The denominator $\binomalal{p}{h} \binomalal{q}{k}$ is the number of terms in the 
above sum.

\medskip
It is also easily verified that the unique symmetric  multiaffine
polar form of degree $\mdeg$
$$f^{\mdeg}_{h, k}((u_{1}, v_{1}), \ldots, (u_{\mdeg}, v_{\mdeg}))$$
of the monomial $u^{h}v^{k}$ is given by
$$f^{\mdeg}_{h, k}((u_{1}, v_{1}), \ldots, (u_{\mdeg}, v_{\mdeg}))
=\frac{1}{\binomalal{m}{h} \binomalal{m-h}{k}}
\sum_{{I \cup J\subseteq\{1,\ldots,m\}}\atop{|I|=h,|J|= k, I \cap J=\emptyset}}
\left(\prod_{i \in I} u_{i}\right)\left(\prod_{j \in J}v_{j}\right).
$$

\medskip
The denominator $\binomalal{m}{h} \binomalal{m-h}{k} = \binomalal{\mdeg}{h\> k\> (\mdeg - h - k)}$ 
is the number of terms in the above sum.

\subsection{Control Points For Polynomial Surfaces}
Let $\Delta_{\mdeg} = \{(i,\, j,\, k)\in\natnums^3
\ |\ i + j + k = \mdeg\}$.
Given an affine frame  $\Delta rst$ in the plane 
(where $r, s, t\in \reals^2$ are affinely
independent points),
a polynomial surface $\mapdef{F}{\affreal^2}{\reals^3}$
of total degree $\mdeg$ specified by the symmetric
multiaffine map
$$\mapdef{f}{(\affreal^2)^{\mdeg}}{\reals^3}$$
is completely determined by the 
family of $\frac{(\mdeg + 1)(\mdeg + 2)}{2}$ points in $\reals^3$ 
$$b_{i,\,j,\,k} = f(\underbrace{r,\ldots,r}_{i}, 
\underbrace{s,\ldots,s}_{j},
\underbrace{t,\ldots,t}_{k}),$$
where $(i,j,k)\in\Delta_{\mdeg}$.

\medskip
These points are called {\it control points\/}, and the family
$\{b_{i,\,j,\,k}\ |\ (i,j,k)\in\Delta_{\mdeg}\}$
is called a {\it triangular control net\/}.

\medskip
Let  $(\ptb{r}_1, \ptb{s}_1)$ and $(\ptb{r}_2, \ptb{s}_2)$ be
any two affine frames for the affine line $\affreal$.
A bipolynomial surface $\mapdef{F}{\affreal^2}{\reals^3}$
of bidegree $\lag \pdeg, \qdeg\rag$ specified by the 
$\pairt{\pdeg}{\qdeg}$-symmetric multiaffine  map
$$\mapdef{f}{\affreal^\pdeg\times \affreal^\qdeg}{\reals^3},$$
is completely determined by the family of $(\pdeg + 1)(\qdeg + 1)$
points in $\reals^3$
$$b_{i,\, j} = f(\underbrace{\ptb{r}_1,\ldots,\ptb{r}_1}_{\pdeg - i}, 
\underbrace{\ptb{s}_1,\ldots,\ptb{s}_1}_{i};
\underbrace{\ptb{r}_2,\ldots,\ptb{r}_2}_{\qdeg - j},
\underbrace{\ptb{s}_2,\ldots,\ptb{s}_2}_{j}),$$
where $0\leq i\leq \pdeg$ and   $0\leq j\leq \qdeg$.

\medskip
These points are called {\it control points\/}, and the family
$\{b_{i,\, j}\ | \ 0\leq i\leq \pdeg,\, 0\leq j\leq \qdeg\}$ 
is called a {\it rectangular control net\/}.

\medskip
Thus, to compute control points, in principle, we need
to compute the polar forms of  polynomials.
However, this method requires  polarization, which is very expensive.
In the following sections, we give recurrence formulae for computing
control points efficiently.
As a corollary, in the case of any affine frame $(r, s)$ or of the
affine frame $((1, 0),\, (0, 1),\, (0, 0))$,
it is possible to give closed-form formulae 
for calculating control points in terms of binomial coefficients.

\section{Computing  Control Points For Curves}
\label{sec5}
We saw in section \ref{sec2} that the polar form of a monomial 
$t^k$ with respect to the degree
$\mdeg$ is
$$f_k^m(t_1,\ldots,t_m) = 
\frac{1}{\binomalal{m}{k}}\,  \sum_{I \subseteq \{1, \ldots, \mdeg\}\atop
            \scriptstyle |I| = k}
         \biggl(\prod_{i\in I} t_i\biggr).$$

\medskip
Letting
$\sigma^{m}_{k} = \binomalal{m}{k}\, f^{\mdeg}_k$,
it is easily verified that we have the following recurrence
equations:

$$\sigma^{m}_{k} =
\cases{       \sigma^{m-1}_{k} + t_m\sigma^{m-1}_{k-1} & if $1\leq k\leq m$;\cr
1 & if $k = 0$ and $m \geq 0$;\cr
       0 & otherwise.\cr
}$$

\medskip
The above formulae can be used to compute inductively the polar values
$$f^{m}_{k}(t_1,\ldots,t_m) =
\frac{1}{\binomalal{m}{k}}\sigma^{m}_{k}(t_1,\ldots,t_m).$$

The computation is reminiscent of the Pascal triangle.
Alternatively, we can compute $f^{m}_{k}$ directly using the recurrence formula
$$f^{m}_{k} = \frac{(\mdeg - k)}{\mdeg}\,f^{m-1}_{k} + 
\frac{k}{\mdeg}t_m\, f^{m-1}_{k-1},$$
where $1\leq k\leq m$.
When writing computer programs implementing these recurrence equations,
we observed that computing $\sigma^{m}_{k}$ and dividing by $\binomalal{m}{k}$ 
is faster than computing $f^{m}_{k}$ directly using the above
formula. This is  because the second method requires
more divisions.

\medskip
Given $(t_1,\ldots,t_\mdeg)$,
computing all the scaled polar values 
$\sigma^{i}_{k}(t_1,\ldots,t_i)$, where
$1\leq k\leq i$ and $1\leq i \leq \mdeg$, requires
time $O(\mdeg^2)$. The naive method using polarization
requires computing $\sum_{i = 0}^{\mdeg}2^{i} = 2^{\mdeg + 1} - 1$ terms.
To compute the coordinates of control points, we simply combine the
$\sigma^{\mdeg}_{k}$. Specifically,
the coordinate value of the control point $b_j$ contributed by the polynomial
$\sum_{k = 0}^{\ndeg} a_kt^k$ 
(where $\ndeg \leq \mdeg$) is
$$\sum_{k = 0}^{\ndeg} a_k\binomalal{\mdeg}{k}^{-1} \sigma^{\mdeg}_{k}
(\underbrace{r,\ldots,r}_{\mdeg - j}, \underbrace{s,\ldots,s}_{j}),$$
where $(r, s)$ is an affine frame.
Given $(t_1,\ldots,t_\mdeg)$,
our algorithm computes the table of 
values $\sigma^{i}_{k}(t_1,\ldots,t_i)$,
where
$1\leq k\leq i$ and $1\leq i \leq \mdeg$,
and thus, it is very cheap to compute these sums.

\medskip
\remark Given a polynomial curve $F$ of degree $\mdeg$
specified by the sequence of control points $(b_0,\ldots,b_\mdeg)$
over $(0, 1)$,
it is well known (see Farin \cite{Farin93,Farin95}, 
Hoschek and Lasser \cite{Hoschek}, or Piegl and Tiller \cite{Piegl})
that $F(t)$ can be expressed in terms of the Bernstein
polynomials $B^\mdeg_k(t) = \binomalal{m}{k}(1 - t)^{m-k} t^k$ as
$$F(t) = B^\mdeg_0(t)\, b_0 + \cdots + B^\mdeg_\mdeg(t)\, b_\mdeg.$$
It is also well known that the Bernstein polynomials
$B^\mdeg_0(t),\ldots$, $B^\mdeg_\mdeg(t)$ form a basis of
the vector space of polynomials of degree $\leq \mdeg$.
Thus, it is also possible to compute the control points
of $F$ by expressing the polynomials involved in the explicit polynomial
definition of $F$ in term of the basis Bernstein polynomials.
Such algorithms were given by
Goldman and Barry \cite{GoldBarry}.
Our algorithm has the same complexity and is more direct.

\medskip
It is also easy to derive closed-form formulae 
for any affine frame  $(r, s)$.

\begin{thm}
If the total degree is $m$ and  there are $r$ occurrences where $t = p$ and $s$ occurrences 
where $t = q$, then
$$f^{\mdeg}_k
 = \frac{\binomalal{p}{k}r^{k}+\binomalal{p}{k-1}\binomalal{q}{1}r^{k-1}s
+\binomalal{p}{k-2}\binomalal{q}{2}r^{k-2}s^{2}+,\ldots,+\binomalal{q}{k}s^{k}}{\binomalal{m}{k}}.$$
\end{thm}

\proof It can be shown by induction using the above
recurrence equations. $\bigsquare$

\section{Computing Rectangular Control Nets}
\label{sec3}
As we saw in  section \ref{sec2},
the polar form of the monomial $u^{h}v^{k}$ with respect to $(p,q)$ is
$$f^{\pdeg, \qdeg}_{h, k}(u_1,\ldots,u_{\pdeg};v_1,\ldots,v_{\qdeg}) 
= \frac{1}{\binomalal{p}{h} \binomalal{q}{k}}\,
\sum_{{I\subseteq\{1,\ldots,p\}, |I|=h}\atop {J\subseteq\{1,\ldots, q\}, |J|= k}}
\left(\prod_{i \in I} u_{i}\right)\left(\prod_{j \in J}v_{j}\right).
$$

\medskip
Letting
$\sigma^{p, q}_{h, k} =
\binomalal{p}{h} \binomalal{q}{k}\, f^{\pdeg, \qdeg}_{h, k}$,
it is easily verified that we have the following recurrence
equations:

$$\sigma^{p, q}_{h, k} =
\cases{       \sigma^{p-1, q-1}_{h, k} + u_p\sigma^{p-1, q-1}_{h-1, k}
+ v_q\sigma^{p-1, q-1}_{h, k-1} + u_pv_q\sigma^{p-1, q-1}_{h-1, k-1} & if $1\leq h \leq p$
and $1 \leq k \leq q$,\cr
 \sigma^{p, q-1}_{0, k} + v_q\sigma^{p, q-1}_{0, k-1} &
if $h = 0\leq p$
and $1 \leq k \leq q$,\cr
\sigma^{p-1, q}_{h, 0} + u_p\sigma^{p-1, q}_{h-1, 0} &
 if $1\leq h \leq p$
and $k = 0 \leq q$,\cr
1 & if $h = k = 0$, $p \geq 0$, and $q\geq 0$;\cr
       0 & otherwise.\cr
}$$

\medskip
Observe that the recurrence formula is a sort of generalization of
the Pascal triangle.
Alternatively, prove that $f^{\pdeg, \qdeg}_{h, k}$ can be computed directly
using the recurrence formula
$$f^{\pdeg, \qdeg}_{h, k} =   
\frac{(\pdeg - h)(\qdeg - k)}{\pdeg\qdeg}\, f^{p-1, q-1}_{h, k} + 
\frac{h(q-k)}{\pdeg\qdeg}u_p\, f^{p-1, q-1}_{h-1, k}
+ \frac{(p-h)k}{\pdeg\qdeg}v_q\, f^{p-1, q-1}_{h, k-1} + 
\frac{hk}{\pdeg\qdeg}u_pv_q\, f^{p-1, q-1}_{h-1, k-1},$$
where  $1\leq h \leq p$ and $1 \leq k \leq q$,
$$f^{\pdeg, \qdeg}_{0, k} =   
\frac{(\qdeg - k)}{\qdeg}\, f^{p, q-1}_{0, k} + 
\frac{k}{\qdeg} v_q\, f^{p, q-1}_{0, k-1},$$
where  $h = 0 \leq p$ and $1 \leq k \leq q$,
and
$$f^{\pdeg, \qdeg}_{h, 0} =   
\frac{(\pdeg - h)}{\pdeg}\, f^{p-1, q}_{h, 0} + 
\frac{h}{\pdeg}u_p\, f^{p-1, q}_{h-1, 0},$$
where  $1 \leq h \leq p$ and $k = 0 \leq q$.
As in section \ref{sec5}, we found that computing $\sigma^{p, q}_{h, k}$
and dividing by $\binomalal{p}{h} \binomalal{q}{k}$
is faster than computing $f^{\pdeg, \qdeg}_{h, k}$ directly.

\medskip
Given $(u_1,\ldots,u_{\pdeg}; v_1,\ldots,v_{\qdeg})$, using the recurrence equations,
computing all  the scaled polar values
$$\sigma^{i, j}_{h, k}(u_1,\ldots, u_{i}; v_1,\ldots,v_{j}),$$
where  $1\leq h \leq i$, $1 \leq k\leq j$, 
$1\leq i \leq \pdeg$, and $1 \leq j\leq \qdeg$,
can be done in time $O(p^2q^2)$. The naive method using polarization
requires computing 
$\sum_{i = 0}^{\pdeg}\sum_{j = 0}^{\qdeg} 2^{i + j} = 2^{\pdeg + \qdeg + 1} - 1$ 
terms.

\medskip
To compute the coordinates of control points, we combine the
$\sigma^{\pdeg, \qdeg}_{h, k}$. Specifically,
the coordinate value of the control point $b_{i, j}$ contributed by the polynomial
$\sum_{h = 0}^{\mdeg}\sum_{k = 0}^{\ndeg} a_{h, k} u^hv^k$ 
(where $\mdeg \leq \pdeg$ and $\ndeg \leq \qdeg$) is
$$\sum_{h = 0}^{\mdeg}\sum_{k = 0}^{\ndeg} a_{h, k} 
\binomalal{\pdeg}{h}^{-1} \binomalal{\qdeg}{k}^{-1} 
\sigma^{\pdeg, \qdeg}_{h, k}
(\underbrace{r_1,\ldots,r_1}_{\pdeg - i}, \underbrace{s_1,\ldots,s_1}_{i};
\underbrace{r_2,\ldots,r_2}_{\qdeg - j}, \underbrace{s_2,\ldots,s_2}_{j}),$$
where $(r_1, s_1)$ and $(r_2, s_2)$ are affine frames.
Our algorithm computes the table of  values
$\sigma^{\mdeg,\ndeg}_{h, k}(u_1,\ldots, u_{\mdeg}; v_1,\ldots,v_{\ndeg})$,
and thus, it is very cheap to compute these sums.

\medskip
When the affine frames $(0, 1)$ are  used,
the following theorem gives closed-form formulae for the polar values
with respect to the bidegree $(p, q)$.

\begin{thm}
If there are $s$ occurrences where  $u = 0$, $r$  occurrences where $v = 0$, 
and all the other occurrences of $u$ and $v$ have the value $1$, then 
$$ f^{\pdeg, \qdeg}_{h,k}(u_{1},u_{2},\ldots,u_{p}; v_{1},v_{2},\ldots,  v_{q})
=\frac{\binomalal{p-s}{h} \binomalal{q-r}{k}}{\binomalal{p}{h} \binomalal{q}{k}}$$.
\end{thm}

\proof It can be shown by induction using the above
recurrence equations. $\bigsquare$

\medskip
It is well known that $F$ can be expressed 
in terms of control points
and (products of) Bernstein polynomials
(see Farin \cite{Farin93,Farin95}, 
Hoschek and Lasser \cite{Hoschek}, or Piegl and Tiller \cite{Piegl}).
As in the case of curves, 
the methods of 
Lodha and Goldman \cite{LodhaGold}
have the same complexity as ours, but our method is more direct.

\section{Computing Triangular Control Nets}
\label{sec4}
As we saw in section \ref{sec2},
the polar form of the monomial $u^{h}v^{k}$ with respect to the total degree $m$ is
$$f^{\mdeg}_{h, k}((u_1, v_1),\ldots, (u_\mdeg, v_\mdeg))
=\frac{1}{\binomalal{m}{h} \binomalal{m-h}{k}}
\sum_{{I \cup J\subseteq\{1,\ldots,m\}}\atop{|I|=h,|J|= k, I \cap J=\emptyset}}
\left(\prod_{i \in I} u_{i}\right)\left(\prod_{j \in J}v_{j}\right).
$$

\medskip
Letting 
$\sigma^{m}_{h, k} = \binomalal{m}{h} \binomalal{m-h}{k}\, f^{\mdeg}_{h,k}$, 
it is easily verified that we have the following recurrence
equations:

$$\sigma^{m}_{h, k} =
\cases{       \sigma^{m-1}_{h, k} + u_m\sigma^{m-1}_{h-1, k}
+ v_m\sigma^{m-1}_{h, k-1}  & if 
$h, k\geq 0$ and $1\leq h + k\leq m$,\cr
1 & if $h = k = 0$ and $m \geq 0$;\cr
       0 & otherwise.\cr
}$$

\medskip
The above formulae can be used to compute inductively the polar values
$$\frac{1}{\binomalal{m}{h} \binomalal{m-h}{k}}
\sigma^{\mdeg}_{h,k}((u_1, v_1),\ldots, (u_\mdeg, v_\mdeg)).$$
The computation consists in building a tetrahedron of values
reminiscent of the Pascal triangle (but $3$-dimensional).
Alternatively, we can compute $f^{m}_{h, k}$ directly using the recurrence formula
$$f^{m}_{h, k} =  \frac{(\mdeg - h - k)}{\mdeg}\,f^{m-1}_{h, k} + 
\frac{h}{\mdeg}u_m\, f^{m-1}_{h-1, k}
+ \frac{k}{\mdeg}v_m\, f^{m-1}_{h, k-1},$$
where $h, k\geq 0$ and $1\leq h + k\leq m$.
Again,  we found that computing $\sigma^{\mdeg}_{h, k}$
and dividing by $\binomalal{m}{h} \binomalal{m-h}{k}$
is faster than computing $f^{\mdeg}_{h, k}$ directly.

\medskip
Given $((u_1, v_1),\ldots,(u_\mdeg, v_{\mdeg}))$, using the recurrence equations,
computing all the scaled polar values
$$\sigma^{i}_{h, k}((u_1, v_1),\ldots, (u_i, v_i)),$$
where $h, k\geq 0$, $1\leq h + k\leq i$, and $1\leq i \leq \mdeg$,
can be done in time $O(\mdeg^3)$.
The naive method using polarization
requires computing 
$\sum_{i=0}^{\mdeg} 3^{i} = (3^{\mdeg + 1} - 1)/2$ 
terms.
To compute the coordinates of control points, we simply combine the
$\sigma^{\mdeg}_{h, k}$. Specifically,
the coordinate value of the control point $b_{i, j, k}$ 
(where $i + j + k = \mdeg$)
contributed by the polynomial
$\sum_{h + l \leq \ndeg} a_{h, l} u^hv^l$ 
(where $\ndeg\leq \mdeg$) is
$$\sum_{h + l \leq \ndeg} a_{h, l} \binomalal{\mdeg}{h\> l\> d}^{-1}
\sigma^{\mdeg}_{h, l}
(\underbrace{r,\ldots,r}_{i}, \underbrace{s,\ldots,s}_{j};
\underbrace{t,\ldots,t}_{k}),$$
where $d = \mdeg - k - l$ and
$r = (r_1, r_2)$, $s = (s_1, s_2)$ and $t = (t_1, t_2)$ are the
vertices of the affine frame.
Our algorithm computes the table of  values
$\sigma^{i}_{h, l}((u_1, v_1),\ldots, (u_i, v_i))$,
and thus, it is very cheap to compute these sums.

\medskip
When the affine frame $((1, 0),\, (0, 1),\, (0, 0))$ is used,
the following theorem gives closed-form formulae for the polar values
with respect to the  total degree $m$.

\begin{thm}
Assume that $ m=r\ +\ s\ +\ t$, with $r$ occurrences of $(1,0)$, 
$s$ occurrences of $ (0,1)$, and $t$ 
occurrences of $(0,0)$  (and no occurrences of $(1,1)$). Then 
$$f^{\mdeg}_{h,k}((u_{1},v_{1}),\ldots, (u_{m}, v_{m}))
=\frac{\binomalal{r}{h}\binomalal{s}{k}}{\binomalal{m}{h}\binomalal{m-h}{k}}$$.
\end{thm}

\proof It can be shown by induction using the above
recurrence equations. $\bigsquare$

\medskip
As in the previous case,
it is well known that $F$ can be expressed 
in terms of control points
and (trivariate) Bernstein polynomials
(see Farin \cite{Farin93,Farin95}, 
Hoschek and Lasser \cite{Hoschek}, or Piegl and Tiller \cite{Piegl}).
The methods of
Lodha and Goldman \cite{LodhaGold}
have the same complexity as ours, but our method is more direct
and very easy to implement.

\section{Examples}
\label{sec6}
We wrote an implementation in {\it Mathematica\/} of a program
computing polar values for curves, using the recurrence equations
of section \ref{sec5}. It works for an arbitrary
affine frame $(r, s)$.

\medskip
As an example, consider the curve of degree $10$ given by
$$\eqaligneno{
x &= \frac{4 t (1 - t^2)^2 (1 - 14 t^2 + t^4)}{(1 + t^2)^5},\cr
y &= \frac{8 t^2 (1 - t^2) (3 - 10 t^2 + 3 t^4)}{(1 + t^2)^5}.\cr  
}$$

Using the above program, the following 
control polygon w.r.t. $[0, 1]$ is obtained:

\begin{verbatim}
rcpoly =  {{0, 0, 1}, {2/5, 0, 1}, {18/25, 12/25, 10/9}, {1/2, 6/5, 4/3},
   {-14/45, 71/45, 12/7}, {-45/37, 45/37, 148/63}, {-71/45, 14/45, 24/7},
   {-6/5, -1/2, 16/3}, {-12/25, -18/25, 80/9}, {0, -2/5, 16}, {0, 0, 32}};
\end{verbatim}

Note that the control points also contain weights, since there are
denominators (see Ramshaw \cite{Ramshaw87}, Farin \cite{Farin93,Farin95}, 
Hoschek and Lasser \cite{Hoschek}, 
or Gallier \cite{Gallbook}).
Here is the  rational curve.

\begin{figure}[H]
\centerline{
\psfig{figure=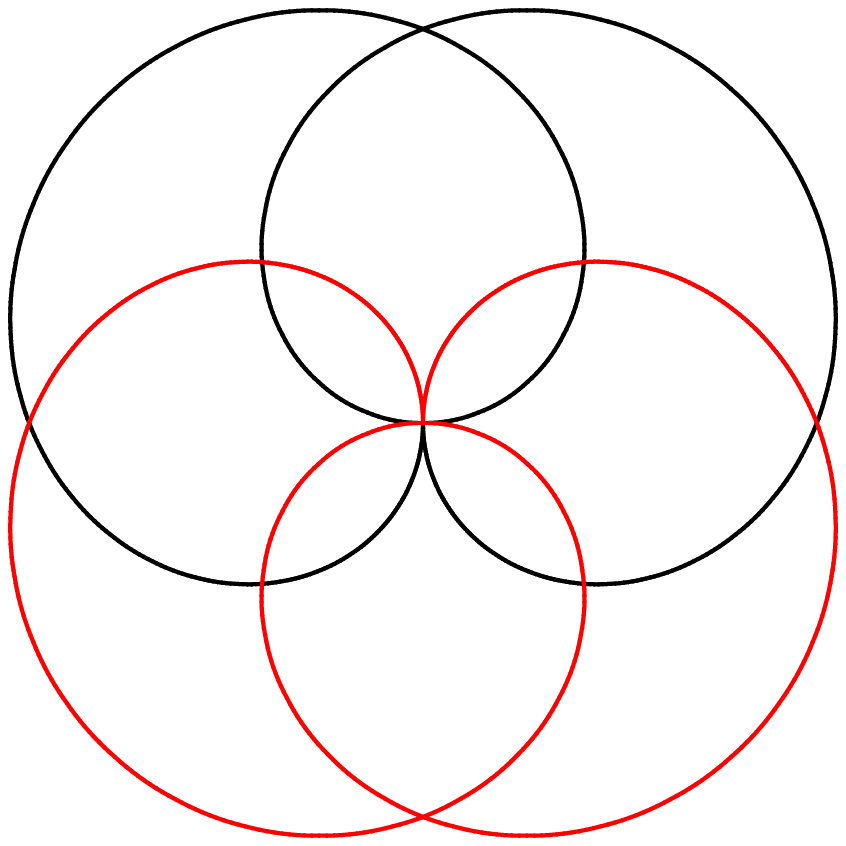,width=4truein}}
\caption{A rose}
\end{figure}

We also wrote an implementation in {\it Mathematica\/} of a program
computing polar values for triangular surface
patches, using the recurrence equations
of section \ref{sec4}. This algorithm works for any affine frame
$(r, s, t)$.

\medskip
In Hilbert and Cohn-Vossen \cite{Hilbert}
(and also do Carmo \cite{DoCarmo76}), an interesting
map $\s{H}$ from $\affreal^3$ to $\affreal^4$ is defined as
$$(x, y, z) \mapsto (xy, yz, xz, x^2 - y^2).$$
This map has the remarkable property that when restricted to the sphere
$S^2$, we have $\s{H}(x, y, z) = \s{H}(x', y', z')$ iff
$(x', y', z') = (x, y, z)$ or
$(x', y', z') = (-x, -y, -z)$. In other words, 
the inverse image of every point in $\s{H}(S^2)$ consists
of two antipodal points.
Thus, the map $\s{H}$ induces an injective map
from the  projective plane onto $\s{H}(S^2)$, which is
obviously continuous, and since the projective plane is
compact, it is a homeomorphism.
Thus, the map $\s{H}$ allows us to realize concretely
the projective plane in $\affreal^4$, by choosing any
parameterization of the sphere $S^2$, and
applying the map $\s{H}$ to it.
For example, the following parametric definition
specifies the entire projective plane over $[-1, 1]\times [-1, 1]$:
$$\eqaligneno{
x &= \frac{16uv^2(1 - u^2)}{(u^2 + 1)^2(v^2 + 1)^2},\cr
y &= \frac{8uv(u^2 + 1)(v^2 - 1)}{(u^2 + 1)^2(v^2 + 1)^2},\cr
z &= \frac{4v(1 - u^4)(v^2 - 1)}{(u^2 + 1)^2(v^2 + 1)^2},\cr
t &= \frac{4v^2(u^4 - 6u^2 + 1)}{(u^2 + 1)^2(v^2 + 1)^2}.\cr
}$$

Using our algorithm,
the following net of degree $8$ over the affine  frame
$((1, 0 ,0)$, $(0, 1, 0)$, $(0, 0, 1))$ is obtained:

\begin{verbatim}
proj8net =  
   {{0, 0, 0, 0, 1}, {0, 0, -1/2, 0, 1}, {0, 0, -14/15, 2/15, 15/14},
   {0, 0, -20/17, 6/17, 17/14}, {0, 0, -120/101, 60/101, 101/70},
   {0, 0, -1, 4/5, 25/14}, {0, 0, -11/16, 15/16, 16/7}, {0, 0, -1/3, 1, 3},
   {0, 0, 0, 1, 4}, {0, 0, 0, 0, 1}, {0, -1/7, -1/2, 0, 1},
   {4/45, -4/15, -14/15, 2/15, 15/14}, {4/17, -28/85, -20/17, 6/17, 17/14},
   {40/101, -32/101, -120/101, 60/101, 101/70},
   {8/15, -6/25, -1, 4/5, 25/14}, {5/8, -1/8, -11/16, 15/16, 16/7},
   {2/3, 0, -1/3, 1, 3}, {0, 0, 0, 0, 15/14}, {0, -4/15, -7/15, 0, 15/14},
   {20/121, -60/121, -105/121, 9/121, 121/105},
   {10/23, -14/23, -25/23, 9/46, 46/35},
   {240/331, -192/331, -360/331, 108/331, 331/210},
   {80/83, -36/83, -75/83, 36/83, 83/42}, {10/9, -2/9, -11/18, 1/2, 18/7},
   {0, 0, 0, 0, 17/14}, {0, -32/85, -7/17, 0, 17/14},
   {9/46, -16/23, -35/46, -1/46, 46/35},
   {27/53, -89/106, -50/53, -3/53, 53/35},
   {36/43, -100/129, -40/43, -4/43, 129/70},
   {12/11, -6/11, -25/33, -4/33, 33/14}, {0, 0, 0, 0, 101/70},
   {0, -48/101, -34/101, 0, 101/70},
   {56/331, -288/331, -204/331, -40/331, 331/210},
   {56/129, -44/43, -98/129, -40/129, 129/70},
   {16/23, -144/161, -120/161, -80/161, 23/10}, {0, 0, 0, 0, 25/14},
   {0, -14/25, -6/25, 0, 25/14}, {8/83, -84/83, -36/83, -16/83, 83/42},
   {8/33, -38/33, -6/11, -16/33, 33/14}, {0, 0, 0, 0, 16/7},
   {0, -5/8, -1/8, 0, 16/7}, {0, -10/9, -2/9, -2/9, 18/7}, {0, 0, 0, 0, 3},
   {0, -2/3, 0, 0, 3}, {0, 0, 0, 0, 4}};
\end{verbatim}

\medskip
Note that the control points also contain weights, since there are
denominators.
If we project the real projective plane onto
a hyperplane in $\affreal^4$, either from a center or parallel to a direction,
we can see a ``3D shadow'' of the real projective plane in $\affreal^3$.
For example, one of the projections is the cross-cap, whose
control net is

\begin{verbatim}
projnet4 = 
   {{0, 0, 0, 1}, {0, 0, 0, 1}, {0, 0, 2/15, 15/14}, {0, 0, 6/17, 17/14},
   {0, 0, 60/101, 101/70}, {0, 0, 4/5, 25/14}, {0, 0, 15/16, 16/7},
   {0, 0, 1, 3}, {0, 0, 1, 4}, {0, 0, 0, 1}, {0, -1/7, 0, 1},
   {4/45, -4/15, 2/15, 15/14}, {4/17, -28/85, 6/17, 17/14},
   {40/101, -32/101, 60/101, 101/70}, {8/15, -6/25, 4/5, 25/14},
   {5/8, -1/8, 15/16, 16/7}, {2/3, 0, 1, 3}, {0, 0, 0, 15/14},
   {0, -4/15, 0, 15/14}, {20/121, -60/121, 9/121, 121/105},
   {10/23, -14/23, 9/46, 46/35}, {240/331, -192/331, 108/331, 331/210},
   {80/83, -36/83, 36/83, 83/42}, {10/9, -2/9, 1/2, 18/7}, {0, 0, 0, 17/14},
   {0, -32/85, 0, 17/14}, {9/46, -16/23, -1/46, 46/35},
   {27/53, -89/106, -3/53, 53/35}, {36/43, -100/129, -4/43, 129/70},
   {12/11, -6/11, -4/33, 33/14}, {0, 0, 0, 101/70}, {0, -48/101, 0, 101/70},
   {56/331, -288/331, -40/331, 331/210}, {56/129, -44/43, -40/129, 129/70},
   {16/23, -144/161, -80/161, 23/10}, {0, 0, 0, 25/14},
   {0, -14/25, 0, 25/14}, {8/83, -84/83, -16/83, 83/42},
   {8/33, -38/33, -16/33, 33/14}, {0, 0, 0, 16/7}, {0, -5/8, 0, 16/7},
   {0, -10/9, -2/9, 18/7}, {0, 0, 0, 3}, {0, -2/3, 0, 3}, {0, 0, 0, 4}}
\end{verbatim}

\medskip
The cross-cap is shown below.

\medskip
\begin{figure}[H]
\centerline{
\epsfig{figure=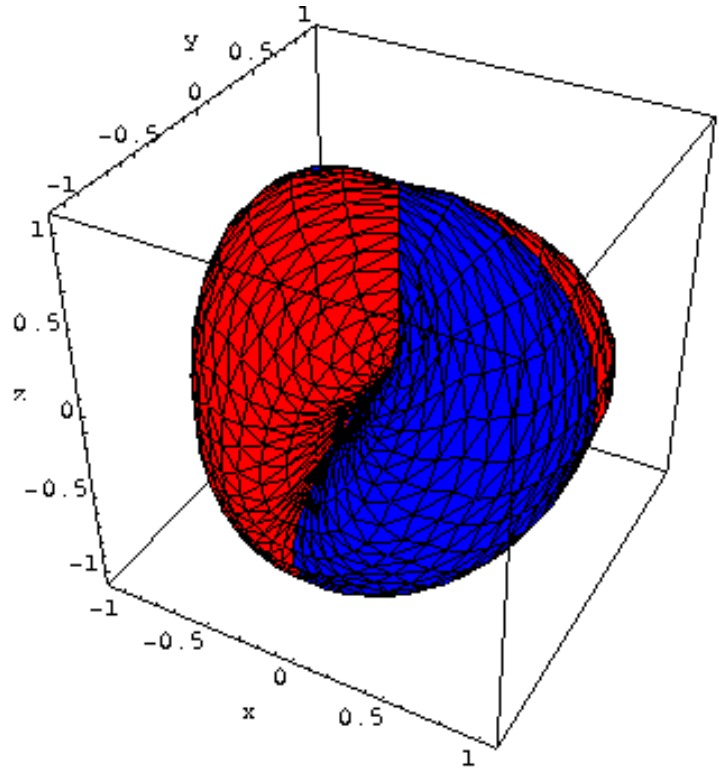,width=4truein}}
\caption{The cross-cap surface}
\end{figure}

\bigskip
{\bf Acknowledgement\/}.
I wish to thank Deepak Tolani for very helpful comments, in particular
on the use of Bernstein polynomials.

\bibliography{../basicmath/cadgeom}

\begin{thebibliography}{10}

\bibitem{DoCarmo76}
Manfredo~P. do~Carmo.
\newblock {\em Differential Geometry of Curves and Surfaces}.
\newblock Prentice Hall, 1976.

\bibitem{Farin95}
Gerald Farin.
\newblock {\em NURB Curves and Surfaces, from Projective Geometry to practical
  use}.
\newblock AK Peters, first edition, 1995.

\bibitem{Farin93}
Gerald Farin.
\newblock {\em Curves and Surfaces for CAGD}.
\newblock Academic Press, fourth edition, 1998.

\bibitem{Gallbook}
Jean~H. Gallier.
\newblock {\em Curves and Surfaces In Geometric Modeling: Theory And
  Algorithms}.
\newblock Morgan Kaufmann, first edition, 1999.

\bibitem{GoldBarry}
Ronald Goldman and P.J. Barry.
\newblock Wonderful triangle: a simple, unified, algorithmic approach to change
  of basis procedures in computer aided geometric design.
\newblock In T.~Lyche and L.L Schumaker, editors, {\em Mathematical Methods in
  CAGD II}, pages 297--320. Academic Press, 1992.

\bibitem{Hilbert}
D.~Hilbert and S.~Cohn-Vossen.
\newblock {\em Geometry and the Imagination}.
\newblock Chelsea Publishing Co., 1952.

\bibitem{Hoschek}
J.~Hoschek and D.~Lasser.
\newblock {\em Computer Aided Geometric Design}.
\newblock AK Peters, first edition, 1993.

\bibitem{LodhaGold}
S.K. Lodha and Ronald Goldman.
\newblock Change of basis algorithms for surfaces in {CAGD}.
\newblock {\em Computer Aided Geometric Design}, 12:801--824, 1995.

\bibitem{Piegl}
Les Piegl and Wayne Tiller.
\newblock {\em The NURBS Book}.
\newblock Monograph in Visual Communications. Springer Verlag, first edition,
  1995.

\bibitem{Ramshaw87}
Lyle Ramshaw.
\newblock Blossoming: A connect-the-dots approach to splines.
\newblock Technical report, Digital SRC, Palo Alto, CA 94301, 1987.
\newblock Report No. 19.

\end{thebibliography}
\bibliographystyle{plain} 
\end{document}